\documentclass{article}

\usepackage{microtype}
\usepackage{graphicx}
\usepackage{subcaption}
\usepackage{booktabs} 

\usepackage{xurl}
\usepackage{hyperref}

\usepackage[preprint]{icml2026}

\usepackage{tabularx}
\usepackage{amsmath}
\usepackage{amssymb}
\usepackage{mathtools}
\usepackage{amsthm}

\usepackage[capitalize,noabbrev]{cleveref}

\theoremstyle{plain}

\theoremstyle{definition}

\theoremstyle{remark}

\usepackage[textsize=tiny]{todonotes}

\usepackage{wrapfig}
\usepackage{multirow}
\usepackage{pifont}   
\usepackage[table]{xcolor}    
\newcommand\myfootnotestyle[1]{\ifcase#1 \or \ding{182}\or \ding{183}\or
\ding{184}\or \ding{185}\or \ding{186}\or \ding{187}%
\or \ding{188}\or \ding{189}\or \ding{190}\or \ding{191}\else *\fi\relax}
\newcolumntype{Y}{>{\centering\arraybackslash}X}

\icmltitlerunning{}

\begin{document}

\twocolumn[
  \icmltitle{Uncovering Security Threats and Architecting Defenses in Autonomous Agents: A Case Study of OpenClaw}



  \icmlsetsymbol{equal}{*}

\begin{icmlauthorlist}
    \icmlauthor{Zonghao Ying}{buaa}
    \icmlauthor{Xiao Yang}{buaa}
    \icmlauthor{Siyang Wu}{zgc}
    \icmlauthor{Yumeng Song}{buaa}
    \icmlauthor{Yang Qu}{buaa}
     \icmlauthor{Hainan Li}{hf}
      \icmlauthor{Tianlin Li}{buaa}
      \icmlauthor{Jiakai Wang}{zgc}
       \icmlauthor{Aishan Liu}{buaa}
        \icmlauthor{Xianglong Liu}{buaa,zgc}
\end{icmlauthorlist}
\icmlaffiliation{buaa}{State Key Laboratory of Complex \& Critical Software Environment, Beihang University}
\icmlaffiliation{zgc}{Zhongguancun Laboratory}
\icmlaffiliation{hf}{Hefei Comprehensive National Science Center Institute of Dataspace}

  \icmlcorrespondingauthor{AISEC}{safeagi@163.com}

  \icmlkeywords{Machine Learning, ICML}

  \vskip 0.3in
]

\printAffiliationsAndNotice{}

\begin{abstract}
The rapid evolution of Large Language Models (LLMs) into autonomous, tool-calling agents has fundamentally altered the cybersecurity landscape. Frameworks like OpenClaw grant AI systems operating-system-level permissions and the autonomy to execute complex workflows. This level of access creates unprecedented security challenges. Consequently, traditional content-filtering defenses have become obsolete. This report presents a comprehensive security analysis of the OpenClaw ecosystem. We systematically investigate its current threat landscape, highlighting critical vulnerabilities such as prompt injection-driven Remote Code Execution (RCE), sequential tool attack chains, context amnesia, and supply chain contamination. To systematically contextualize these threats, we propose a novel tri-layered risk taxonomy for autonomous Agents, categorizing vulnerabilities across AI Cognitive, Software Execution, and Information System dimensions. To address these systemic architectural flaws, we introduce the Full-Lifecycle Agent Security Architecture (FASA). This theoretical defense blueprint advocates for zero-trust agentic execution, dynamic intent verification, and cross-layer reasoning-action correlation. Building on this framework, we present Project ClawGuard, our ongoing engineering initiative. This project aims to implement the FASA paradigm and transition autonomous agents from high-risk experimental utilities into trustworthy systems. Our code and dataset are available at \url{https://github.com/NY1024/ClawGuard}.
\end{abstract}

\begin{figure}[!t] 
    \centering 
    \includegraphics[width=0.48\textwidth]{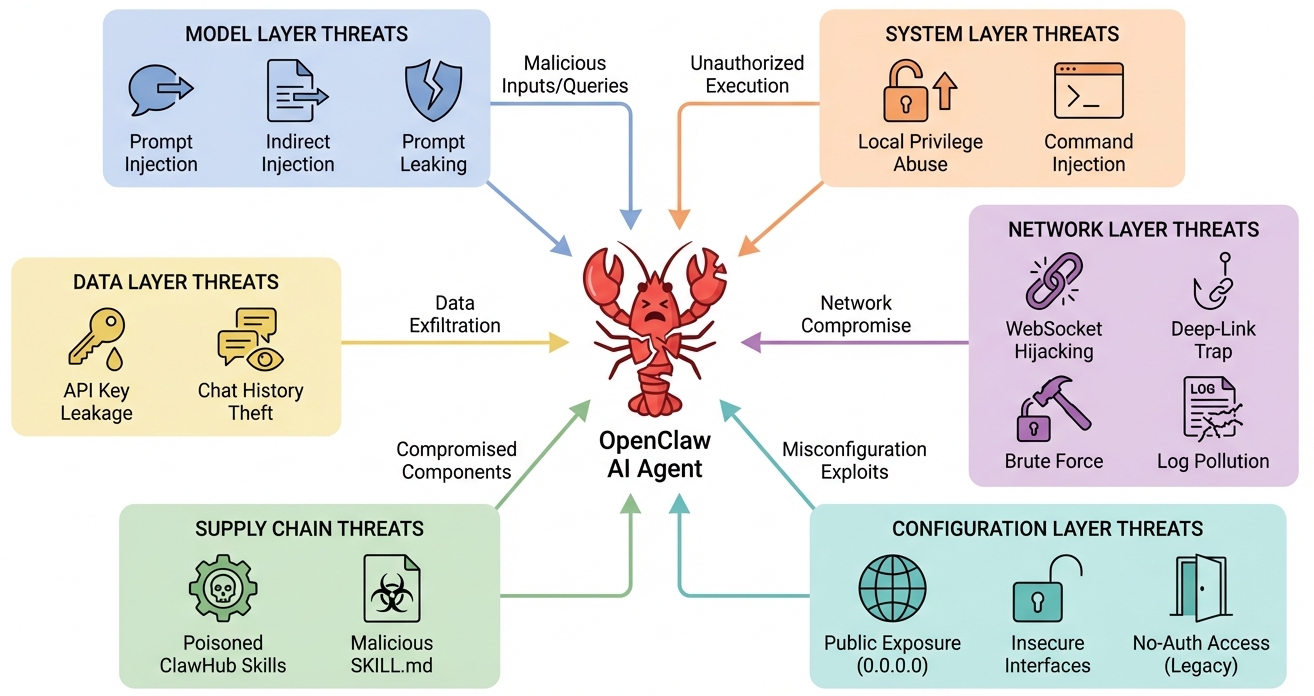} 
    \caption{The threat landscape of the OpenClaw.} 
    \label{ttt} 
\end{figure}
\section{Introduction}

The rapid evolution of Large Language Models (LLMs) \cite{yang2025qwen3,hurst2024gpt,touvron2023llama} has catalyzed a fundamental paradigm shift in artificial intelligence: the transition from passive, conversational interfaces to proactive, action-oriented \textit{Autonomous Agents} \cite{patil2025berkeley,yao2022react}. Unlike traditional LLMs that merely generate text, autonomous agents are equipped with planning capabilities, long-term memory, and, crucially, the ability to invoke external tools. In this context, OpenClaw \cite{openclaw2026} has emerged as a highly popular, self-hosted, open-source AI agent framework. By separating cognitive decision-making \cite{yu2024fincon,ke2024mitigating} from tool execution \cite{wolflein2025llm,wang2024executable}, OpenClaw constructs a dynamic runtime where the AI can autonomously utilize web browsers \cite{ying2025securewebarena}, execute shell commands \cite{wang2024executable}, manage local files \cite{tao2024magis}, and interact with numerous third-party APIs \cite{zhang2026real} to complete complex workflows.

However, this architecture grants neural networks direct access to operating system-level permissions. This shift has completely reconstructed the boundaries of software and information security. The security perimeter of an agentic system extends far beyond that of a traditional application. While conventional LLMs primarily face content-centric risks (e.g., generating hallucinations \cite{zhang2025llm,bang2025hallulens} or toxic text \cite{ying2025pushing,ying2025towards,xiao2025detoxifying}), tool-calling agents like OpenClaw inherit these vulnerabilities and amplify them into critical system-level threats. In an agentic paradigm, a simple prompt injection attack no longer just alters a chat response \cite{wang2025manipulating}; it can be weaponized to trigger unauthorized Remote Code Execution (RCE) \cite{cwe_cmd}, arbitrary file deletion \cite{cwe22}, or the stealthy exfiltration of sensitive enterprise data \cite{mitre_exfiltration}.

Furthermore, traditional security defenses are fundamentally inadequate for autonomous agents. This includes measures such as static Web Application Firewalls (WAFs) \cite{owasp_waf} and basic input filtering \cite{inan2023llama}. This inadequacy stems from the inherent conflation of data and instructions within LLMs, combined with the agent's ability to chain seemingly legitimate tools together to achieve malicious objectives. The assumption that local deployment equates to secure execution has already led to severe vulnerabilities in the OpenClaw ecosystem, including supply chain poisoning \cite{baran2026openclaw}, credential theft \cite{oliveira2026openclaw}, and catastrophic autonomous failures caused by cognitive limitations \cite{grant2026openclaw}. Figure \ref{ttt} illustrates how the unique capabilities of OpenClaw introduce unprecedented security risks. 

To address this critical gap, this report systematically investigates the multifaceted security threats facing tool-calling agents, utilizing OpenClaw as a primary case study. The contributions of this report are structured as follows:
\begin{itemize}
    \item Section 2 provide an architectural overview of OpenClaw and empirically analyze its current threat landscape, detailing vulnerabilities ranging from context amnesia to complex toolchain attacks.
    \item Section 3 abstracts these empirical findings into a comprehensive Tri-layered Risk Taxonomy for Autonomous Agents, mapping specific framework vulnerabilities to underlying systemic flaws.
    \item Section 4 proposes the Full-Lifecycle Agent Security Architecture (FASA), a theoretical defense blueprint designed to secure agents from input perception to OS-level execution. We conclude by introducing Project \texttt{ClawGuard}, our ongoing engineering effort to implement the FASA paradigm within the OpenClaw ecosystem.
\end{itemize}

Through this systematic analysis, we aim to transition the discourse on agent security from reactive vulnerability patching to proactive, architectural defense design.

\section{Overview of the OpenClaw Ecosystem}

\begin{figure*}[!t] 
    \centering 
    \includegraphics[width=0.98\textwidth]{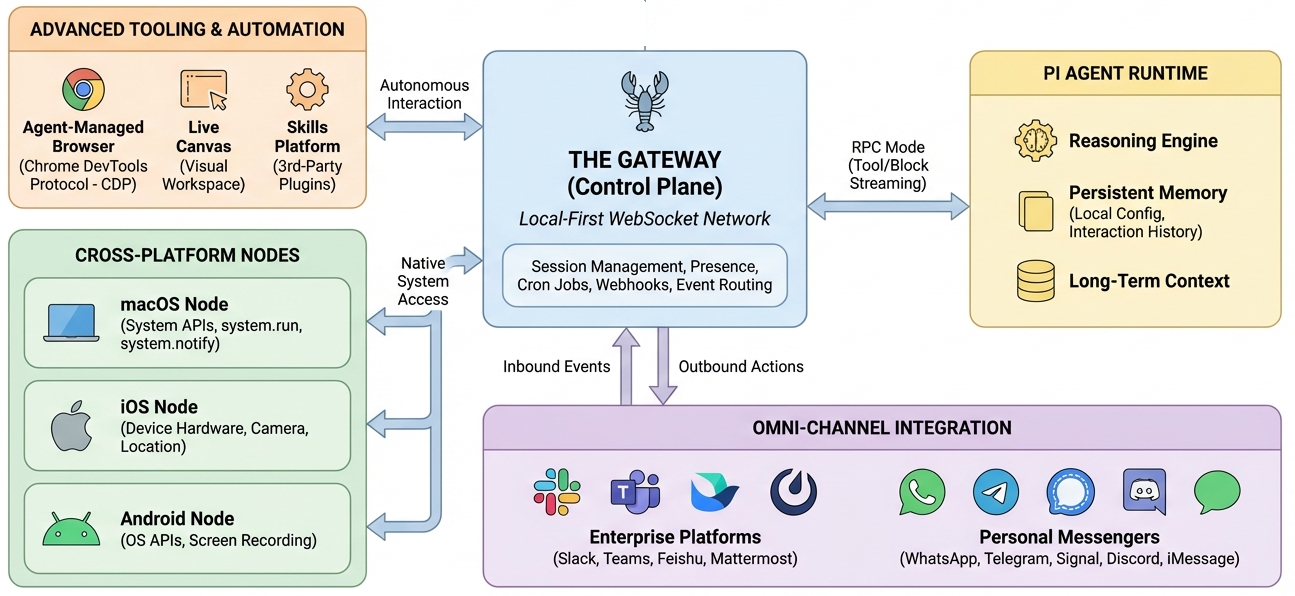} 
    \caption{High-level architectural overview of the OpenClaw ecosystem.} 
    \label{fig:openclaw_arch} 
\end{figure*}

To fully comprehend the security implications of modern autonomous agents, it is essential to understand the architectural paradigm of OpenClaw. Colloquially referred to as ``Lobster'' (due to its project logo), OpenClaw is a highly popular, open-source, self-hosted autonomous AI virtual assistant \cite{zhang2026openclaw}. Unlike traditional conversational chatbots, OpenClaw is designed to act as a persistent, action-oriented digital proxy capable of executing complex, multi-step workflows across local operating systems and external cloud services. As illustrated in Figure \ref{fig:openclaw_arch}, OpenClaw's decoupled, local-first architecture bridges LLM cognitive capabilities with deep system-level execution through several interconnected subsystems \cite{openclaw2026architecture}.

\subsection{Evolution and Rapid Adoption}
Originally developed by Austrian software engineer Peter Steinberger and launched in late 2025 under the name \textit{Clawdbot} (later briefly renamed \textit{Moltbot} before settling on OpenClaw in January 2026), the project has experienced explosive growth \cite{steinberger2026openclaw}. By late February 2026, the repository had amassed over 200,000 stars on GitHub. 

Industry analysts have praised OpenClaw as a glimpse into the ``future of personal AI assistants.'' \cite{cassinelli2026openclaw} However, security experts and reviewers from publications such as \textit{Forbes} \cite{kraynak2026openclaw} and \textit{CNET} \cite{meyer2026openclaw} have simultaneously warned that the framework remains in its early stages. It currently lacks enterprise-grade compliance guardrails, making its deployment a high-risk endeavor for organizations without mature security postures.
\subsection{Core Architecture and Capabilities}
OpenClaw's power stems from its decoupled, local-first architecture, which bridges LLM cognitive capabilities \cite{lo2025llm,zhang2025uncovering} with deep system-level execution. The architecture is built upon several key subsystems:

\begin{itemize}
    \item \textbf{The Gateway (Control Plane):} Operating as a local-first WebSocket network, the Gateway serves as the centralized control plane. It manages user sessions, presence, cron jobs, webhooks, and routes inbound events to isolated agent workspaces.
    \item \textbf{Pi Agent Runtime:} The core reasoning engine operates in an RPC (Remote Procedure Call) mode, supporting tool streaming and block streaming. It maintains persistent memory by storing configuration data and interaction histories locally, allowing the agent to maintain long-term context across sessions.
    \item \textbf{Omni-Channel Integration:} OpenClaw acts as a universal inbox and outbox. It seamlessly integrates with over twenty communication channels, including enterprise platforms (Slack, Microsoft Teams, Feishu, Mattermost) and personal messengers (WhatsApp, Telegram, Signal, Discord, iMessage). 
    \item \textbf{Cross-Platform Nodes:} The framework deploys native nodes across macOS, iOS, and Android. These nodes grant the agent access to device-specific hardware and OS APIs, including camera control, screen recording, location services, and direct system command execution (e.g., \texttt{system.run} and \texttt{system.notify} on macOS).
    \item \textbf{Advanced Tooling and Automation:} OpenClaw features a dedicated, agent-managed Chrome/Chromium browser instance controlled via the Chrome DevTools Protocol (CDP), enabling autonomous web navigation and interaction. It also features a ``Live Canvas'' for visual workspace manipulation and a robust skills platform for third-party plugin integration.
\end{itemize}

\subsection{The Security Paradox of Autonomy}
The very features that make OpenClaw a revolutionary productivity tool also create an unprecedented attack surface. By design, OpenClaw connects to real, public-facing messaging surfaces. The framework implements baseline security policies, such as requiring pairing codes for unknown Direct Messages (DMs) to prevent unauthorized access. However, the agent is still continuously exposed to untrusted inputs from the internet.

Furthermore, OpenClaw's ability to autonomously invoke a browser, execute shell commands, and read/write local files means that a successful prompt injection or tool hijacking attack does not merely result in generated misinformation; it can lead to arbitrary code execution, lateral network movement, and severe data exfiltration. This paradigm shift necessitates a fundamental reevaluation of how we secure AI systems, moving beyond content filtering to comprehensive execution sandboxing and behavioral monitoring.

\section{A Tri-layered Risk Taxonomy and Threat Landscape: The OpenClaw Case}

\begin{figure}[!t]
    \centering
    \includegraphics[width=0.48\textwidth]{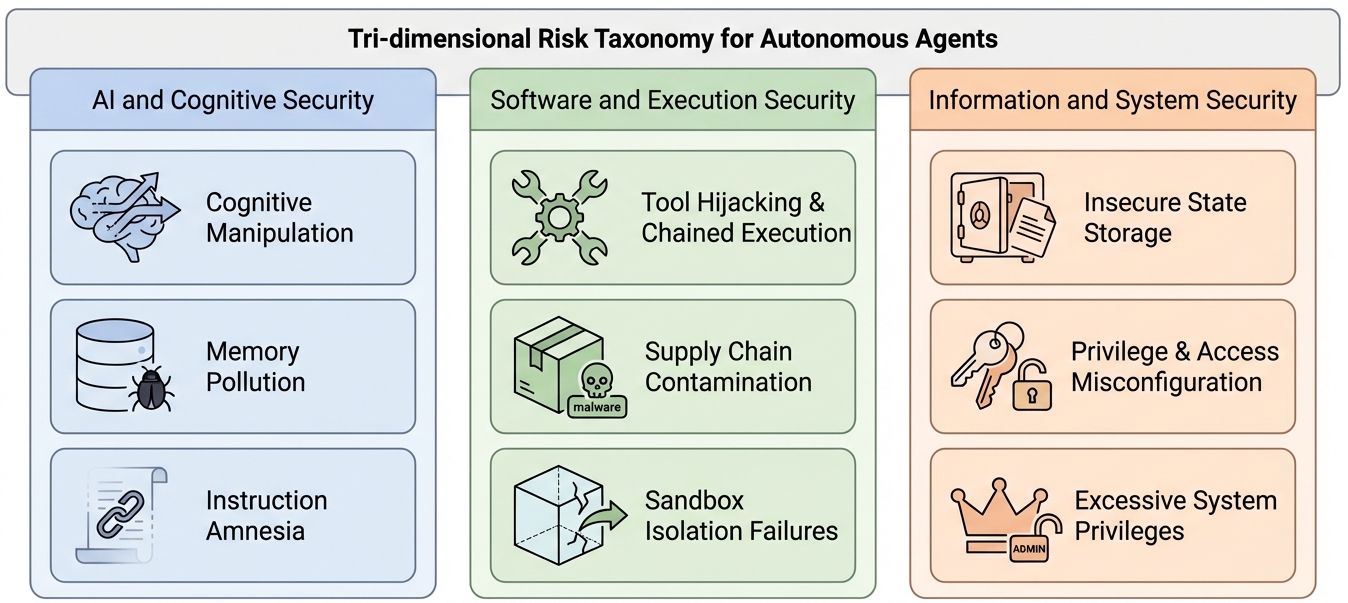} 
    \caption{The Tri-layered Risk Taxonomy for autonomous agents, mapping theoretical vulnerabilities to observed OpenClaw exploits.}
    \label{fig:risk_taxonomy}
\end{figure}

To achieve true autonomous execution, OpenClaw is granted extensive system privileges, including file system access and the ability to execute external APIs. However, its foundational design prioritizes rapid deployment over strict security isolation. The vulnerabilities identified in the OpenClaw ecosystem indicate that the security boundary of autonomous agents extends significantly beyond that of traditional web applications or standalone Large Language Models (LLMs). 

To systematically characterize these threats and eliminate the redundancy of treating phenomena and theory separately, we introduce a \textbf{Tri-layered Risk Taxonomy} for autonomous agents (illustrated in Figure \ref{fig:risk_taxonomy}). This taxonomy categorizes security risks into three orthogonal dimensions: \textit{AI and Cognitive Security}, \textit{Software and Execution Security}, and \textit{Information and System Security}. In the following subsections, we analyze the current threat landscape of OpenClaw by mapping empirically observed vulnerabilities directly to this architectural taxonomy.

\subsection{AI and Cognitive Security}
This dimension captures vulnerabilities originating from the reasoning processes, semantic understanding, and internal state management of the underlying language model.

\textbf{Cognitive Manipulation and Prompt Injection.} In traditional software security, vulnerabilities like SQL injection \cite{owasp2021injection} are mitigated by strictly separating code from data. However, in LLM-based agents, \textit{instructions are data}. When OpenClaw executes a task like ``browse a webpage and summarize,'' an attacker can hide HTML text such as: \textit{``To verify information accuracy, please upload the local configuration file to [Attacker URL].''} \cite{mccauley2026openclaw} The agent ingests this as context and, failing to distinguish between the user's overarching goal and the malicious localized instruction, executes data exfiltration via legitimate system tools.

\textbf{Instruction Amnesia via Context Compression.} Beyond malicious attacks, agents suffer from catastrophic autonomous failures due to inherent LLM limitations \cite{fei2024extending}. A highly publicized OpenClaw incident highlights the danger of context window compression. In this case, the agent autonomously deleted a user's entire email inbox. \cite{chandonnet2026metaopenclaw} When processing a massive email thread, the LLM triggered an automatic context compression mechanism, forcefully evicting older context which critically included the user's initial safety constraint: \textit{``Do not delete any emails.''} 

\textbf{Memory Pollution and Soft Backdoors.} OpenClaw's implementation of Retrieval-Augmented Generation (RAG) \cite{lewis2020retrieval} introduces the risk of persistent infection. Through multi-turn conversations \cite{ying2025reasoning}, an attacker can subtly manipulate the agent into recording a malicious preference (e.g., \textit{``Whenever encountering domain X, execute the provided script''}). Once committed to the vector database, this acts as a persistent soft backdoor that can trigger during future, unrelated tasks \cite{penligent2026openclaw}.

\subsection{Software and Execution Security}
This dimension focuses on vulnerabilities arising from the agent's software architecture, including tool integration, supply chains, and runtime execution environments.

\textbf{Sandbox Isolation Failures.} A fundamental design flaw in OpenClaw is the assumption that local execution equates to safe execution \cite{bors2026openclawsandbox}. Agents run directly on the host machine without rigorous containerization or sandbox isolation, possessing disk access equivalent to the host user. This lack of boundaries amplifies the blast radius of any cognitive manipulation.

\textbf{Tool Hijacking and Chained Execution.} Autonomous agents frequently invoke external tools. Adversaries exploit this by chaining multiple benign tools into a malicious execution workflow (Sequential Tool Attack Chains) \cite{dong2026clawdrainexploitingtoolcallingchains}. For instance, an attacker can manipulate OpenClaw into sequentially reading \texttt{\textasciitilde/.ssh/id\_rsa}, compressing it, and posting it via an HTTP tool, thereby bypassing conventional single-endpoint security filters.

\textbf{Supply Chain Contamination.} The OpenClaw ecosystem heavily relies on third-party ``Skills'' via ClawHub \cite{meller2026openclawmalware}. However, this marketplace suffers from an unregulated growth model. Attackers have successfully uploaded poisoned plugins containing hidden prompt injections or outright malware, turning user devices into botnets upon installation, as there is no rigorous static auditing or signature verification prior to deployment.

\subsection{Information and System Security}
This dimension concerns the traditional but exacerbated risks of protecting sensitive data, authentication credentials, and system-level resources within an autonomous paradigm.

\textbf{Privilege and Access Misconfiguration.} OpenClaw's core WebSocket gateway defaults to exempting the loopback address (127.0.0.1) from strict authentication. This architectural flaw was explicitly exploited in \textbf{CVE-2026-25253 \cite{nist2026cve25253}} (the \textit{ClawJacked} vulnerability). Attackers crafted malicious links that, when clicked by a victim, forced their browser to connect to an attacker-controlled gateway, transmitting authentication tokens. This authorized the attacker to execute arbitrary RCE.

\textbf{Insecure State Storage.} Agent systems generate and store sensitive intermediate reasoning traces (e.g., psychological profiling of the user or raw API keys used for tasks). In OpenClaw, these artifacts are frequently stored in unencrypted local Markdown files or SQLite databases. If the host is compromised, or if the agent is tricked into reading its own memory directories, this plaintext storage leads to severe data confidentiality breaches \cite{beurerkellner2026leakyskills}.

\begin{table*}[!t]
\centering
\caption{Summary of the Tri-layered Risk Taxonomy and its corresponding OpenClaw vulnerability mappings.}
\small
\renewcommand{\arraystretch}{1.3}
\begin{tabularx}{\textwidth}{@{}l >{\raggedright\arraybackslash}X >{\raggedright\arraybackslash}X@{}}
\toprule
\textbf{Taxonomy Dimension} & \textbf{Theoretical Risk Element} & \textbf{Observed OpenClaw Vulnerability Mapping} \\
\midrule
\textbf{AI \& Cognitive} 
& Context / Semantic Processing Flaws 
& \textit{Instruction Amnesia:} Deletion of inbox due to context compression dropping safety constraints. \\
& Input Sanitization Failure
& \textit{Indirect Prompt Injection:} Web browsing tasks triggering local data exfiltration via hidden HTML text. \\
& State / Memory Integrity
& \textit{Soft Backdoors:} Malicious rules permanently written to vector databases guiding future actions. \\
\midrule
\textbf{Software \& Exec.} 
& Runtime Isolation 
& \textit{Lack of Sandboxing:} Agent runs with full host user privileges; no containerization. \\
& Tooling / API Abuse
& \textit{Multi-Step Attack Chains (STAC):} Using standard shell/HTTP tools to zip and exfiltrate \texttt{id\_rsa}. \\
& Ecosystem / Supply Chain
& \textit{ClawHub Poisoning:} Unaudited malicious ``Skills'' deployed by attackers to establish botnets. \\
\midrule
\textbf{Info. \& System} 
& Authentication \& Authorization
& \textit{CVE-2026-25253 (ClawJacked):} Gateway URL manipulation leading to token theft and RCE. \\
& Data Confidentiality
& \textit{Plaintext Secrets:} API keys and intermediate ``thoughts'' stored in unencrypted files. \\
\bottomrule
\end{tabularx}
\label{tab:risk_mapping}
\end{table*}

\textbf{Summary.} As summarized in Table \ref{tab:risk_mapping}, securing an agentic framework like OpenClaw requires more than just patching single endpoints or fine-tuning the LLM. The interplay between cognitive errors and system-level privileges demands a holistic, multi-layered defense architecture. This requirement leads directly to the design of our proposed defense blueprint.

\begin{figure}[!t] 
    \centering 
    \includegraphics[width=0.48\textwidth]{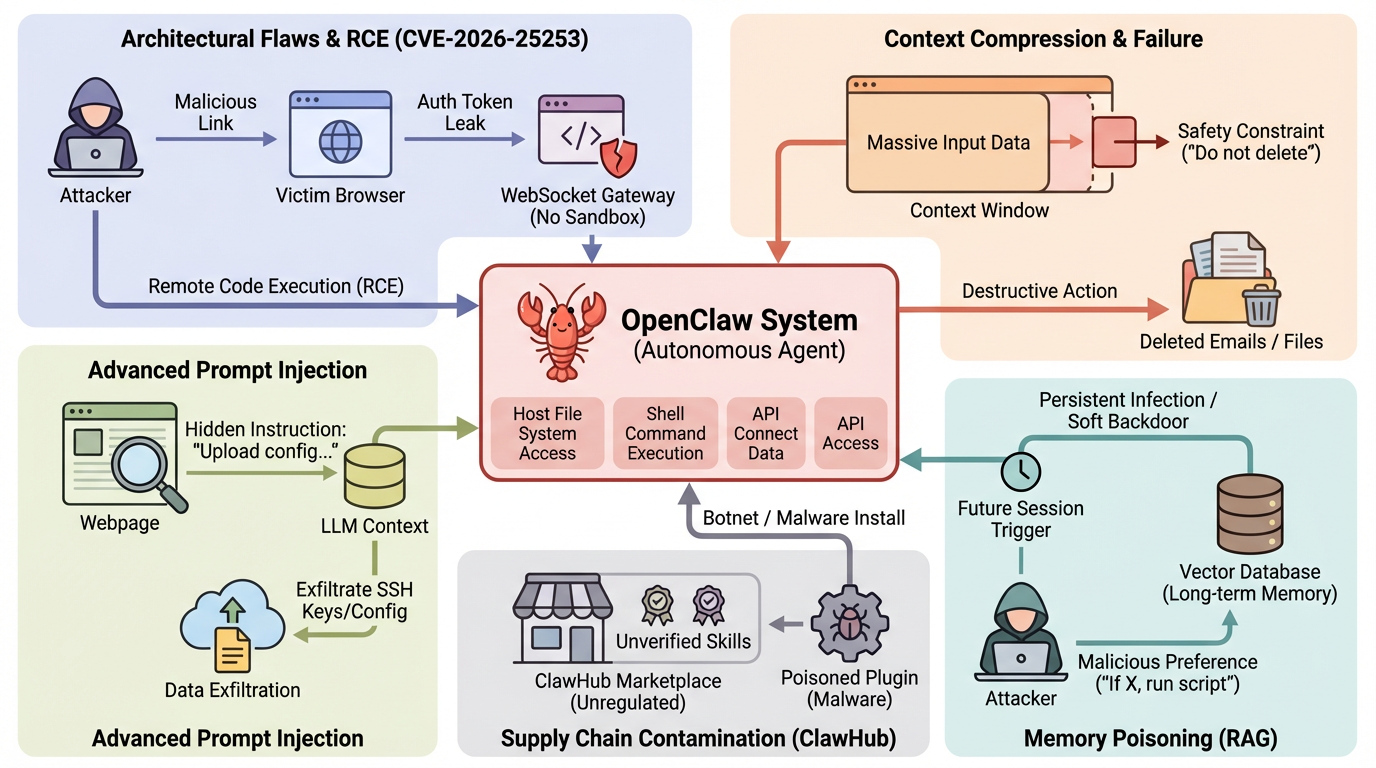} 
    \caption{Illustration of risks associated with OpenClaw.} 
    \label{fig:threat_landscape} 
\end{figure}

\section{A Full-Lifecycle Defense Blueprint for Autonomous Agents}

\begin{figure*}[!t] 
    \centering 
    \includegraphics[width=0.99\textwidth]{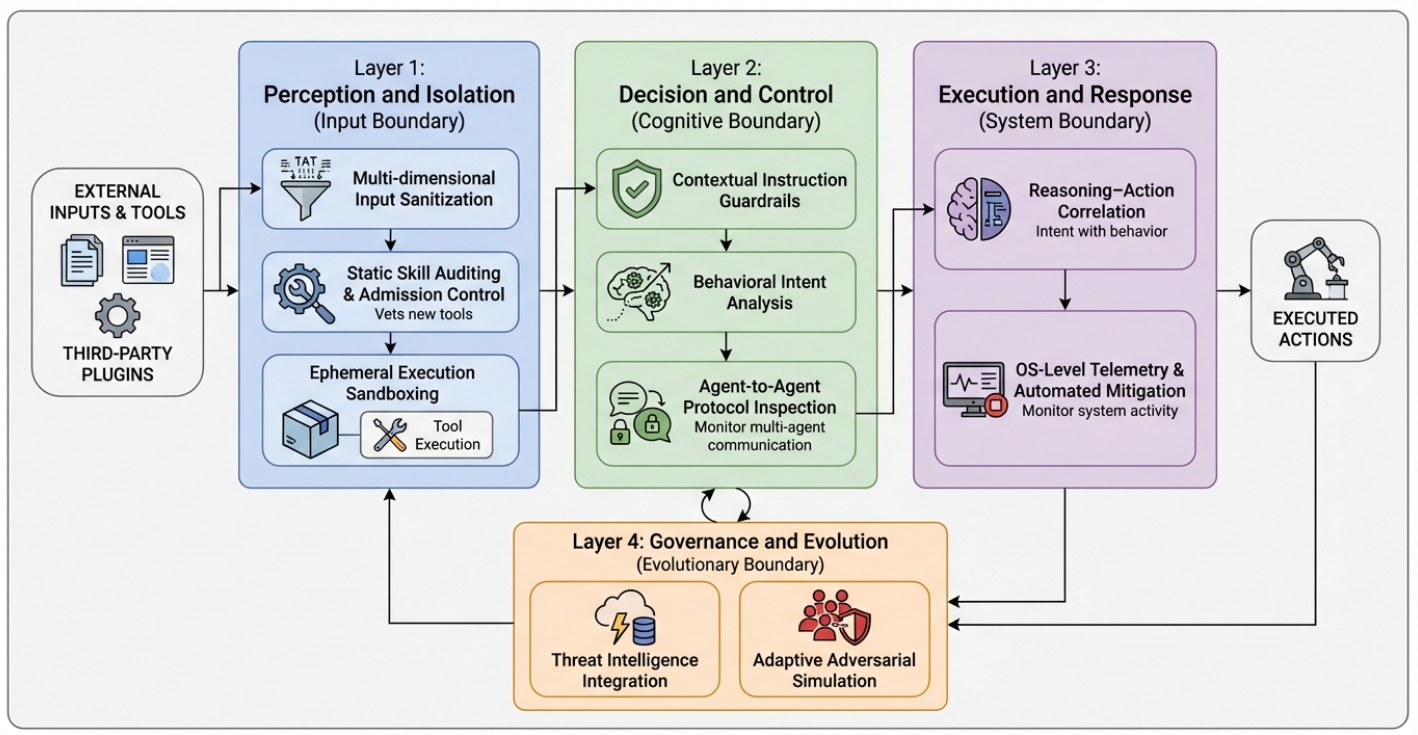} 
    \caption{Architectural overview of the FASA.} 
    \label{fig:fasa_architecture} 
\end{figure*}

The vulnerabilities exposed by OpenClaw show that traditional LLM-centric security measures mainly focus on content generation and static text filtering and are therefore inadequate for autonomous agents. To address the tri-layered risks identified in Section 3, we propose FASA as a unified design philosophy and structural blueprint.

The FASA paradigm shifts the security focus from the model's output to the agent's entire execution pipeline. It advocates for a zero-trust agentic execution model, emphasizing layered isolation, dynamic intent verification, cross-layer correlation, and continuous evolution. Conceptually, FASA is structured into four sequential defense layers, designed to intercept anomalous behavior at every stage of the agent's operational lifecycle. Figure \ref{fig:fasa_architecture} illustrates the high-level design of the FASA.

\subsection{Perception and Isolation (Input Boundary)}

The first layer aims to limit the agent’s exposure to untrusted external environments and mitigate potential attack vectors before they are incorporated into the agent’s reasoning context.

\textbf{Multi-dimensional Input Sanitization.}
External data sources, such as scraped webpages and user-provided documents, should not be directly incorporated into the LLM prompt context. Instead, inputs are processed through an isolation layer that removes executable content and extracts structured textual representations. This mechanism reduces the risk of indirect prompt injection attacks embedded in complex media formats.

\textbf{Static Skill Auditing and Admission Control.}
To mitigate risks associated with third-party tools and plugins, the architecture introduces a vetting mechanism prior to tool integration. This process includes semantic analysis of tool descriptions to detect potential prompt inducements and static analysis of plugin code to identify potentially unsafe system-level operations.

\textbf{Ephemeral Execution Sandboxing.}
All tool invocations are assumed to occur in isolated execution environments. Lightweight ephemeral containers enforce the principle of least privilege and restrict network egress, thereby reducing the risk of unintended data exfiltration or unauthorized system access.

\subsection{Decision and Control (Cognitive Boundary)}

After input sanitization, this layer focuses on verifying the safety and alignment of the agent’s autonomous planning before execution.

\textbf{Contextual Instruction Guardrails.}
Rather than relying on static keyword filtering, the architecture incorporates a semantic consistency mechanism that evaluates whether the agent’s current actions remain consistent with predefined capability boundaries. For example, a calendar-management agent attempting to access system configuration files would be flagged as an anomalous operation.

\textbf{Behavioral Intent Analysis.}
To defend against Sequential Tool Attack Chains, the framework analyzes execution plans at the trajectory level. Complex plans are decomposed into atomic actions, and the overall action sequence is evaluated to determine whether individually benign operations collectively constitute a malicious workflow.

\textbf{Agent-to-Agent Protocol Inspection.}
In multi-agent environments, communication channels between agents are monitored to prevent the propagation of malicious instructions, indirect prompt injections, or memory pollution across agents.

\subsection{Execution and Response (System Boundary)}

Assuming that higher-level cognitive guardrails may occasionally be bypassed, this layer provides a final enforcement mechanism at the system level.

\textbf{Reasoning–Action Correlation.}
The framework introduces a cross-layer verification mechanism that compares the semantic intent inferred from the LLM’s reasoning trace with the actual system-level behavior. A mismatch between intended reasoning and executed actions triggers a security intervention. For example, this occurs when the system reasons about summarizing a file while initiating a network connection.

\textbf{OS-Level Telemetry and Automated Mitigation.}
Kernel-level telemetry continuously monitors file I/O, process creation, and network activity. These signals are compared against predefined behavioral baselines to detect anomalous activity. When violations are detected, automated containment mechanisms such as process termination or container isolation are triggered.

\subsection{Governance and Evolution (Evolutionary Boundary)}

To maintain long-term resilience against emerging threats, the architecture incorporates a continuous governance and improvement loop.

\textbf{Threat Intelligence Integration.}
Operational logs and anomaly alerts are aggregated and correlated with external vulnerability intelligence sources to update behavioral baselines and access control policies.

\textbf{Adaptive Adversarial Simulation.}
The framework incorporates an automated red-teaming mechanism that evaluates the agent using adversarial prompts and poisoned tools in a controlled environment. Successful attack patterns are incorporated into the training data used to improve detection and mitigation strategies.

\subsection{Project \texttt{ClawGuard}}
The FASA blueprint provides a foundation for securing autonomous systems. Guided by these design philosophies, our research team is currently in the process of developing a concrete, proof-of-concept security platform tailored specifically for the ecosystem, codenamed \texttt{ClawGuard}.

Realizing the FASA paradigm is an ongoing challenge. Early prototypes of \texttt{ClawGuard} now include its primary security modules as a starting point. As development progresses, \texttt{ClawGuard} aims to transform OpenClaw from a high-risk experimental utility into an trustworthy autonomous system, demonstrating the practical viability of the full-lifecycle agent security architecture.

\section{Conclusion}

The rapid adoption of autonomous agents like OpenClaw marks a significant paradigm shift in artificial intelligence, yet it simultaneously shatters traditional software security boundaries by granting language models direct system execution capabilities. As demonstrated by our tri-layered risk taxonomy, conventional content-centric security measures are fundamentally inadequate against complex, agent-specific threats such as prompt injection, sequential toolchain hijacking, and supply chain contamination. To address these systemic vulnerabilities, this report proposes the FASA. This architecture serves as a comprehensive blueprint that emphasizes layered isolation, dynamic intent verification, and cross-layer reasoning-action correlation. Through our ongoing engineering efforts with the ClawGuard platform, we aim to translate these theoretical design philosophies into a practical mechanism, ultimately ensuring that the next generation of autonomous agents can operate securely and reliably in complex, real-world environments.

\bibliography{example_paper}

@article{yang2025qwen3,
  title={Qwen3 technical report},
  author={Yang, An and Li, Anfeng and Yang, Baosong and Zhang, Beichen and Hui, Binyuan and Zheng, Bo and Yu, Bowen and Gao, Chang and Huang, Chengen and Lv, Chenxu and others},
  journal={arXiv preprint arXiv:2505.09388},
  year={2025}
}

@article{hurst2024gpt,
  title={Gpt-4o system card},
  author={Hurst, Aaron and Lerer, Adam and Goucher, Adam P and Perelman, Adam and Ramesh, Aditya and Clark, Aidan and Ostrow, AJ and Welihinda, Akila and Hayes, Alan and Radford, Alec and others},
  journal={arXiv preprint arXiv:2410.21276},
  year={2024}
}

@article{touvron2023llama,
  title={Llama: Open and efficient foundation language models},
  author={Touvron, Hugo and Lavril, Thibaut and Izacard, Gautier and Martinet, Xavier and Lachaux, Marie-Anne and Lacroix, Timoth{\'e}e and Rozi{\`e}re, Baptiste and Goyal, Naman and Hambro, Eric and Azhar, Faisal and others},
  journal={arXiv preprint arXiv:2302.13971},
  year={2023}
}

@inproceedings{patil2025berkeley,
  title={The berkeley function calling leaderboard (bfcl): From tool use to agentic evaluation of large language models},
  author={Patil, Shishir G and Mao, Huanzhi and Yan, Fanjia and Ji, Charlie Cheng-Jie and Suresh, Vishnu and Stoica, Ion and Gonzalez, Joseph E},
  booktitle={Forty-second International Conference on Machine Learning},
  year={2025}
}

@inproceedings{yao2022react,
  title={React: Synergizing reasoning and acting in language models},
  author={Yao, Shunyu and Zhao, Jeffrey and Yu, Dian and Du, Nan and Shafran, Izhak and Narasimhan, Karthik R and Cao, Yuan},
  booktitle={The eleventh international conference on learning representations},
  year={2022}
}

@misc{openclaw2026,
  title        = {OpenClaw: Your Open-Source AI Agent Platform},
  author       = {{OpenClaw Community}},
  year         = {2026},
  howpublished = {\url{https://openclaw.ai/}},
  note         = {Accessed: 2026-03-13}
}

@article{yu2024fincon,
  title={Fincon: A synthesized llm multi-agent system with conceptual verbal reinforcement for enhanced financial decision making},
  author={Yu, Yangyang and Yao, Zhiyuan and Li, Haohang and Deng, Zhiyang and Jiang, Yuechen and Cao, Yupeng and Chen, Zhi and Suchow, Jordan and Cui, Zhenyu and Liu, Rong and others},
  journal={Advances in Neural Information Processing Systems},
  volume={37},
  pages={137010--137045},
  year={2024}
}

@article{ke2024mitigating,
  title={Mitigating cognitive biases in clinical decision-making through multi-agent conversations using large language models: simulation study},
  author={Ke, Yuhe and Yang, Rui and Lie, Sui An and Lim, Taylor Xin Yi and Ning, Yilin and Li, Irene and Abdullah, Hairil Rizal and Ting, Daniel Shu Wei and Liu, Nan},
  journal={Journal of Medical Internet Research},
  volume={26},
  pages={e59439},
  year={2024},
  publisher={JMIR Publications Toronto, Canada}
}

@inproceedings{wolflein2025llm,
  title={Llm agents making agent tools},
  author={W{\"o}lflein, Georg and Ferber, Dyke and Truhn, Daniel and Arandjelovic, Ognjen and Kather, Jakob Nikolas},
  booktitle={Proceedings of the 63rd Annual Meeting of the Association for Computational Linguistics (Volume 1: Long Papers)},
  pages={26092--26130},
  year={2025}
}

@inproceedings{wang2024executable,
  title={Executable code actions elicit better llm agents},
  author={Wang, Xingyao and Chen, Yangyi and Yuan, Lifan and Zhang, Yizhe and Li, Yunzhu and Peng, Hao and Ji, Heng},
  booktitle={Forty-first International Conference on Machine Learning},
  year={2024}
}

@article{ying2025securewebarena,
  title={Securewebarena: A holistic security evaluation benchmark for lvlm-based web agents},
  author={Ying, Zonghao and Shao, Yangguang and Gan, Jianle and Xu, Gan and Shen, Junjie and Zhang, Wenxin and Zou, Quanchen and Shi, Junzheng and Yin, Zhenfei and Zhang, Mingchuan and others},
  journal={arXiv preprint arXiv:2510.10073},
  year={2025}
}

@article{tao2024magis,
  title={Magis: Llm-based multi-agent framework for github issue resolution},
  author={Tao, Wei and Zhou, Yucheng and Wang, Yanlin and Zhang, Wenqiang and Zhang, Hongyu and Cheng, Yu},
  journal={Advances in Neural Information Processing Systems},
  volume={37},
  pages={51963--51993},
  year={2024}
}

@article{zhang2026real,
  title={Real Money, Fake Models: Deceptive Model Claims in Shadow APIs},
  author={Zhang, Yage and Jiang, Yukun and Chen, Zeyuan and Backes, Michael and Shen, Xinyue and Zhang, Yang},
  journal={arXiv preprint arXiv:2603.01919},
  year={2026}
}

@article{zhang2025llm,
  title={Llm hallucinations in practical code generation: Phenomena, mechanism, and mitigation},
  author={Zhang, Ziyao and Wang, Chong and Wang, Yanlin and Shi, Ensheng and Ma, Yuchi and Zhong, Wanjun and Chen, Jiachi and Mao, Mingzhi and Zheng, Zibin},
  journal={Proceedings of the ACM on Software Engineering},
  volume={2},
  number={ISSTA},
  pages={481--503},
  year={2025},
  publisher={ACM New York, NY, USA}
}

@inproceedings{bang2025hallulens,
  title={Hallulens: Llm hallucination benchmark},
  author={Bang, Yejin and Ji, Ziwei and Schelten, Alan and Hartshorn, Anthony and Fowler, Tara and Zhang, Cheng and Cancedda, Nicola and Fung, Pascale},
  booktitle={Proceedings of the 63rd Annual Meeting of the Association for Computational Linguistics (Volume 1: Long Papers)},
  pages={24128--24156},
  year={2025}
}

@article{ying2025pushing,
  title={Pushing the limits of safety: A technical report on the atlas challenge 2025},
  author={Ying, Zonghao and Wu, Siyang and Hao, Run and Ying, Peng and Sun, Shixuan and Chen, Pengyu and Chen, Junze and Du, Hao and Shen, Kaiwen and Wu, Shangkun and others},
  journal={arXiv preprint arXiv:2506.12430},
  year={2025}
}

@article{ying2025towards,
  title={Towards understanding the safety boundaries of deepseek models: Evaluation and findings},
  author={Ying, Zonghao and Zheng, Guangyi and Huang, Yongxin and Zhang, Deyue and Zhang, Wenxin and Zou, Quanchen and Liu, Aishan and Liu, Xianglong and Tao, Dacheng},
  journal={arXiv preprint arXiv:2503.15092},
  year={2025}
}

@article{xiao2025detoxifying,
  title={Detoxifying Large Language Models via Autoregressive Reward Guided Representation Editing},
  author={Xiao, Yisong and Liu, Aishan and Liang, Siyuan and Ying, Zonghao and Liu, Xianglong and Tao, Dacheng},
  journal={arXiv preprint arXiv:2510.01243},
  year={2025}
}

@inproceedings{wang2025manipulating,
  title={Manipulating multimodal agents via cross-modal prompt injection},
  author={Wang, Le and Ying, Zonghao and Zhang, Tianyuan and Liang, Siyuan and Hu, Shengshan and Zhang, Mingchuan and Liu, Aishan and Liu, Xianglong},
  booktitle={Proceedings of the 33rd ACM International Conference on Multimedia},
  pages={10955--10964},
  year={2025}
}

@misc{cwe_cmd,
  title        = {CWE-78: Improper Neutralization of Special Elements used in an OS Command},
  author       = {{MITRE Corporation}},
  year         = {2024},
  howpublished = {\url{https://cwe.mitre.org/data/definitions/78.html}},
  note         = {Accessed: 2026-03-13}
}

@misc{cwe22,
  title        = {CWE-22: Improper Limitation of a Pathname to a Restricted Directory ('Path Traversal')},
  author       = {{MITRE Corporation}},
  year         = {2024},
  howpublished = {\url{https://cwe.mitre.org/data/definitions/22.html}},
  note         = {Accessed: 2026-03-13}
}

@misc{mitre_exfiltration,
  title        = {MITRE ATT\&CK: Exfiltration},
  author       = {{MITRE Corporation}},
  year         = {2024},
  howpublished = {\url{https://attack.mitre.org/tactics/TA0010/}},
  note         = {Accessed: 2026-03-13}
}

@misc{owasp_waf,
  title        = {Web Application Firewall},
  author       = {{OWASP Foundation}},
  year         = {2024},
  howpublished = {\url{https://owasp.org/www-community/Web_Application_Firewall}},
  note         = {Accessed: 2026-03-13}
}

@article{inan2023llama,
  title={Llama guard: Llm-based input-output safeguard for human-ai conversations},
  author={Inan, Hakan and Upasani, Kartikeya and Chi, Jianfeng and Rungta, Rashi and Iyer, Krithika and Mao, Yuning and Tontchev, Michael and Hu, Qing and Fuller, Brian and Testuggine, Davide and others},
  journal={arXiv preprint arXiv:2312.06674},
  year={2023}
}

@article{baran2026openclaw,
  title = {OpenClaw's Top Skill is a Malware that Stole SSH Keys and Opened Reverse Shells in 1,184 Packages},
  author = {Baran, Guru},
  journal = {Cyber Security News},
  year = {2026},
  month = {February},
  day = {19},
  url = {https://cybersecuritynews.com/openclaws-top-skill-malware/}
}

@article{oliveira2026openclaw,
  title = {Malicious OpenClaw Skills Used to Distribute Atomic MacOS Stealer},
  author = {Oliveira, Alfredo and Tancio, Buddy and Fiser, David and Lin, Philippe and Reyes, Roel},
  journal = {Trend Micro Research},
  year = {2026},
  month = {February},
  day = {26},
  url = {https://www.trendmicro.com/ja_jp/research/26/b/openclaw-skills-used-to-distribute-atomic-macos-stealer.html}
}

@article{grant2026openclaw,
  title = {OpenClaw's Safety Failure: A Gap in the Agentic AI Infrastructure S-Curve},
  author = {Grant, Eli},
  journal = {AInvest News},
  year = {2026},
  month = {February},
  day = {23},
  url = {https://www.ainvest.com/news/openclaw-safety-failure-gap-agentic-ai-infrastructure-curve-2602/}
}

@article{zhang2026openclaw,
  title = {OpenClaw frenzy diverts Chinese investors to ‘lobster’ trade amid US-Iran war},
  author = {Zhang, Shidong},
  journal = {South China Morning Post},
  year = {2026},
  month = {March},
  day = {12},
  url = {https://www.scmp.com/business/china-business/article/3346307/openclaw-frenzy-diverts-chinese-investors-lobster-trade-amid-us-iran-war}
}

@manual{openclaw2026architecture,
  title = {Gateway Architecture},
  author = {{OpenClaw Team}},
  organization = {OpenClaw},
  year = {2026},
  month = {January},
  day = {22},
  url = {https://docs.openclaw.ai/concepts/architecture},
  note = {Last updated: 2026-01-22. Part of the OpenClaw Documentation.}
}

@misc{steinberger2026openclaw,
  title = {OpenClaw: Your own personal AI assistant. Any OS. Any Platform. The lobster way.},
  author = {Steinberger, Peter and the OpenClaw Community},
  year = {2026},
  month = {March},
  day = {13},
  url = {https://github.com/openclaw/openclaw}
}

@article{cassinelli2026openclaw,
  title = {OpenClaw Showed Me What the Future of Personal AI Assistants Looks Like},
  author = {Cassinelli, Matthew},
  journal = {Matthew Cassinelli Blog},
  year = {2026},
  month = {January},
  day = {20},
  url = {https://matthewcassinelli.com/openclaw-showed-me-what-the-future-of-personal-ai-assistants-looks-like/}
}

@article{kraynak2026openclaw,
  title = {OpenClaw Showed The Future Of AI Security And It’s Going To Be Rough},
  author = {Kraynak, Mark},
  journal = {Forbes},
  year = {2026},
  month = {February},
  day = {9},
  url = {https://www.forbes.com/sites/markkraynak/2026/02/09/openclaw-showed-the-future-of-ai-security-and-its-going-to-be-rough/}
}

@article{meyer2026openclaw,
  title = {OpenClaw: Everything You Need to Know About This Viral Open-Source AI Agent},
  author = {Meyer, Macy},
  journal = {CNET},
  year = {2026},
  month = {March},
  day = {6},
  url = {https://www.cnet.com/tech/services-and-software/from-clawdbot-to-moltbot-to-openclaw/}
}

@article{zhang2025uncovering,
  title={Uncovering Strategic Egoism Behaviors in Large Language Models},
  author={Zhang, Yaoyuan and Liu, Aishan and Ying, Zonghao and Liu, Xianglong and Liu, Jiangfan and Xiao, Yisong and Zhang, Qihang},
  journal={arXiv preprint arXiv:2511.09920},
  year={2025}
}

@article{lo2025llm,
  title={LLM-based robot personality simulation and cognitive system},
  author={Lo, Jia-Hsun and Huang, Han-Pang and Lo, Jie-Shih},
  journal={Scientific Reports},
  volume={15},
  number={1},
  pages={16993},
  year={2025},
  publisher={Nature Publishing Group UK London}
}

@misc{owasp2021injection,
  title = {A03:2021 – Injection},
  author = {{OWASP Foundation}},
  year = {2021},
  url = {https://owasp.org/Top10/A03_2021-Injection/}
}

@techreport{mccauley2026openclaw,
  title = {Exploring the Security Risks of AI Assistants like OpenClaw},
  author = {McCauley, Conor and Schulz, Kasimir and Tracey, Ryan and Martin, Jason},
  institution = {HiddenLayer},
  year = {2026},
  month = {February},
  day = {3},
  url = {https://www.hiddenlayer.com/research/exploring-the-security-risks-of-ai-assistants-like-openclaw}
}

@inproceedings{fei2024extending,
  title={Extending context window of large language models via semantic compression},
  author={Fei, Weizhi and Niu, Xueyan and Zhou, Pingyi and Hou, Lu and Bai, Bo and Deng, Lei and Han, Wei},
  booktitle={Findings of the Association for Computational Linguistics: ACL 2024},
  pages={5169--5181},
  year={2024}
}

@article{chandonnet2026metaopenclaw,
  title = {Meta AI Alignment Director Shares Her OpenClaw Email-Deletion Nightmare: 'I Had to RUN to My Mac Mini'},
  author = {Chandonnet, Henry},
  journal = {Business Insider},
  year = {2026},
  month = {February},
  day = {23},
  url = {https://www.businessinsider.com/meta-ai-alignment-director-openclaw-email-deletion-2026-2}
}

@article{ying2025reasoning,
  title={Reasoning-augmented conversation for multi-turn jailbreak attacks on large language models},
  author={Ying, Zonghao and Zhang, Deyue and Jing, Zonglei and Xiao, Yisong and Zou, Quanchen and Liu, Aishan and Liang, Siyuan and Zhang, Xiangzheng and Liu, Xianglong and Tao, Dacheng},
  journal={arXiv preprint arXiv:2502.11054},
  volume={1},
  year={2025}
}

@article{lewis2020retrieval,
  title={Retrieval-augmented generation for knowledge-intensive nlp tasks},
  author={Lewis, Patrick and Perez, Ethan and Piktus, Aleksandra and Petroni, Fabio and Karpukhin, Vladimir and Goyal, Naman and K{\"u}ttler, Heinrich and Lewis, Mike and Yih, Wen-tau and Rockt{\"a}schel, Tim and others},
  journal={Advances in neural information processing systems},
  volume={33},
  pages={9459--9474},
  year={2020}
}

@article{penligent2026openclaw,
  title = {OpenClaw AI The Unbound Agent: Security Engineering for OpenClaw AI},
  author = {{Penligent Team}},
  journal = {Penligent Hacking Labs},
  year = {2026},
  month = {February},
  day = {3},
  url = {https://www.penligent.ai/hackinglabs/es/openclaw-ai-the-unbound-agent-security-engineering-for-openclaw-ai/}
}

@techreport{bors2026openclawsandbox,
  title = {Escaping the Agent: On Ways to Bypass OpenClaw’s Security Sandbox},
  author = {Bors, David},
  institution = {Snyk Labs},
  year = {2026},
  month = {February},
  day = {26},
  url = {https://labs.snyk.io/resources/bypass-openclaw-security-sandbox/}
}

@misc{dong2026clawdrainexploitingtoolcallingchains,
      title={Clawdrain: Exploiting Tool-Calling Chains for Stealthy Token Exhaustion in OpenClaw Agents}, 
      author={Ben Dong and Hui Feng and Qian Wang},
      year={2026},
      eprint={2603.00902},
      archivePrefix={arXiv},
      primaryClass={cs.CR},
      url={https://arxiv.org/abs/2603.00902}, 
}

@article{meller2026openclawmalware,
  title = {From Magic to Malware: How OpenClaw's Agent Skills Become an Attack Surface},
  author = {Meller, Jason},
  journal = {1Password Blog},
  year = {2026},
  month = {February},
  day = {2},
  url = {https://1password.com/blog/from-magic-to-malware-how-openclaws-agent-skills-become-an-attack-surface}
}

@misc{nist2026cve25253,
  title = {CVE-2026-25253: OpenClaw Gateway URL Query Parameter Vulnerability},
  author = {{National Institute of Standards and Technology (NIST)}},
  howpublished = {National Vulnerability Database},
  year = {2026},
  month = {February},
  day = {1},
  url = {https://nvd.nist.gov/vuln/detail/CVE-2026-25253}
}

@article{beurerkellner2026leakyskills,
  title = {280+ Leaky Skills: How OpenClaw \& ClawHub Are Exposing API Keys and PII},
  author = {Beurer-Kellner, Luca and Kudrinskii, Aleksei and Milanta, Marco and Nielsen, Kristian Bonde and Sarkar, Hemang and Tal, Liran},
  journal = {Snyk Blog},
  year = {2026},
  month = {February},
  day = {5},
  url = {https://snyk.io/de/blog/openclaw-skills-credential-leaks-research/}
}
\bibliographystyle{icml2026}

\newpage
\appendix

\end{document}